# TOWARDS A FASTER SYMBOLIC AGGREGATE APPROXIMATION METHOD

Muhammad Marwan Muhammad Fuad[1], Pierre-François Marteau[1]
[1]*VALORIA, Université de Bretagne Sud, Université Européenne de Bretagne*
*BP. 573, 56017 Vannes, France*
*{marwan.fuad, pierre-francois.marteau}@univ-ubs.fr*

Abstract: The similarity search problem is one of the main problems in time series data mining. Traditionally, this problem was tackled by sequentially comparing the given query against all the time series in the database, and returning all the time series that are within a predetermined threshold of that query. But the large size and the high dimensionality of time series databases that are in use nowadays make that scenario inefficient. There are many representation techniques that aim at reducing the dimensionality of time series so that the search can be handled faster at a lower-dimensional space level. The symbolic aggregate approximation (SAX) is one of the most competitive methods in the literature. In this paper we present a new method that improves the performance of SAX by adding to it another exclusion condition that increases the exclusion power. This method is based on using two representations of the time series: one of SAX and the other is based on an optimal approximation of the time series. Pre-computed distances are calculated and stored offline to be used online to exclude a wide range of the search space using two exclusion conditions. We conduct experiments which show that the new method is faster than SAX.

## 1 INTRODUCTION

Similarity search is a hot topic in computer science. This problem has many applications in multimedia databases, bioinformatics, pattern recognition, text mining, computer vision, and others. Small structured databases can be handled easily. But managing large unstructured, or weakly structured, databases requires serious effort, especially when they contain complex data types.

Time series are data types that appear in many applications. Time series data mining includes many tasks such as classification, clustering, similarity search, motif discovery, anomaly detection, and others. Research in time series data mining has focussed on two aspects; the first aspect is the dimensionality reduction techniques that can represent the time series efficiently and effectively at lower-dimensional spaces. Different indexing structures are used to handle time series.

Time series are high dimensional data , so even indexing structures can fail in handling these data because of what is known as the "dimensionality curse" phenomenon. One of the best solutions to deal with this phenomenon is to utilize a dimensionality reduction method to reduce the dimensionality of the time series, then to use a suitable indexing structure on the reduced space.

There have been different suggestions to represent time series in lower dimensional spaces. To mention a few; *Discrete Fourier Transform* (DFT) (Agrawal *et al* . 1993) and (Agrawal *et al* . 1995), *Discrete Wavelet Transform* (DWT) (Chan and Fu 1999), *Singular Value Decomposition* (SVD) (Korn *et al* . 1997), *Adaptive Piecewise Constant Approximation* (APCA) (Keogh *et al* . 2001), *Piecewise Aggregate Approximation* (PAA) (Keogh *et al* . 2000) and ( Yi and Faloutsos 2000), *Piecewise Linear Approximation* (PLA)

(Morinaka et al . 2001), *Chebyshev Polynomials* (CP) (Cai and Ng 2004).

Among the different representation methods, symbolic representation has several advantages, because it allows researchers to benefit from text-retrieval algorithms and techniques that are widely used in the text mining and bioinformatics communities (Keogh et al . 2001).

The other aspect of research in time series data mining is similarity distances. There are quite a large number of similarity distances; some of them are applied to a particular data type, while others can be applied to different data types.

Among the different similarity distances, there are those that can be used on symbolic data types. At first they were available for data types whose representation is naturally symbolic (DNA and proteins sequences, textual data…etc). But later these symbolic similarity distances were extended to apply to other data types that can be transformed into sequences by using an appropriate symbolic representation technique.

In time series data mining there are several symbolic representation methods. Of all these methods, the symbolic aggregate approximation method (SAX) (Jessica et al . 2003) stands out as one of the most powerful methods. The main feature of this method is that the similarity distance that it uses is easy to compute, because it uses statistical lookup tables.

In this paper we present a new method that speeds up SAX by adding a new exclusion condition that increases the exclusion power of SAX. Our method keeps the original features of SAX, mainly its speed, since the new exclusion condition is pre-computed offline.

The rest of this paper is organized as follows: in section 2 we present background on dimensionality reduction, and on symbolic representation of time series in general, and SAX in particular. The proposed method is presented in section 3. In section 4 we present some of the results of the different experiments we conducted. The conclusion is presented in section 5.

## 2 BACKGROUND

### 2.1 Representation Methods

Managing high-dimensional data is a difficult problem in time series data mining. Representation methods try to overcome this problem by embedding the time series of the original space into a lower dimensional space. Time series are highly correlated data, so representation methods that aim at reducing dimensionality by projecting the original data onto lower dimensional spaces and processing the query in those reduced spaces is a scheme that is widely used in time series data mining community.

When embedding the original space into a lower dimensional space and performing the similarity query in the transformed space, two main side-effects may be encountered; *false alarms*, also called *false positivity,* and *false dismissals*. False alarms are data objects that belong to the response set in the transformed space, but do not belong to the response set in the original space. False dismissals are data objects that the search algorithm excluded in the transformed space, although they are answers to the query in the original space. Generally, false alarms are more tolerated than false dismissals, because a post-processing scan is usually performed on the results of the query in the transformed space to filter out these data objects that are not valid answers to the query in the original space. However, false alarms can slow down the search time if the algorithm returns too many of them. False dismissals are a more serious problem and they need more sophisticated procedures to avoid them.

False alarms and false dismissals are dependent on the transformation used in the embedding. If $f$ is a transformation from the original space $(S_{orig}, d_{orig})$ into another space $(S_{trans}, d_{trans})$ then in order to guarantee no false dismissals this transform should satisfy:

$$d_{trans}(f(u_1), f(u_2)) \le d_{orig}(u_1, u_2),$$
$$\forall u_1, u_2 \in S_{orig} \quad (1)$$

The above condition is known as the *lower-bounding lemma.* ( Yi and Faloutsos 2000)

If a transformation can make the two above distances equal for all the data objects in the original space, then similarity search produces no false alarms or false dismissals. Unfortunately, such an ideal transformation is very hard to find. Yet, we try to make the above distances as close as possible.

The above condition can be written as:

$$0 \le \frac{d_{trans}(f(u_1), f(u_2))}{d_{orig}(u_1, u_2)} \le 1 \quad (2)$$

A *tight* transformation is one that makes the above ratio as close as possible to 1.

## 2.2 SAX

Symbolic representation of time series has attracted much attention recently, because by using this method we can not only reduce the dimensionality of time series, but also benefit from the numerous algorithms used in bioinformatics and text data mining. However, first symbolic representation methods were ad hoc and did not give satisfactory results. But later more sophisticated methods emerged. Of all these method, the symbolic aggregate approximation method (SAX) is one of the most powerful ones in time series data mining.

SAX is based on the fact that normalized time series have highly Gaussian distribution (Larsen and Marx 1986), so by determining the breakpoints that correspond to the chosen alphabet size, one can obtain equal sized areas under the Gaussian curve.

SAX is applied in the following steps: in the first step all the time series in the database are normalized. In the second step, the dimensionality of the time series is reduced by using the PAA. In PAA the times series is divided into equal sized segments or "frames" and the mean value of the points that lie within the frame is computed. The lower dimensional vector of the original time series is the vector whose components are the means of all successive frames of the time series. In the third step, the PAA representation of the time series is discretized. This is achieved by determining the number and the location of the breakpoints. This number is related to the desired alphabet size (which is chosen by the user), i.e.

*alphabet_size=number(breakpoints)+1* . Their locations are determined by statistical lookup tables, so that these breakpoints produce equal-sized areas under the Gaussian curve. The interval between two successive breakpoints is assigned to a symbol of the alphabet, and each segment of the PAA that lies within that interval is discretized by that symbol.

The last step of SAX is using the following similarity distance:

$$MINDIST(\widetilde{S},\widetilde{T}) \equiv \sqrt{\frac{n}{N}} \sqrt{\sum_{i=1}^{N}(dist(\widetilde{s_i},\widetilde{t_i}))^2} \quad (3)$$

Where $n$ is the length of the original time series, $N$ is the number of the frames, $\widetilde{S}$ and $\widetilde{T}$ are the symbolic

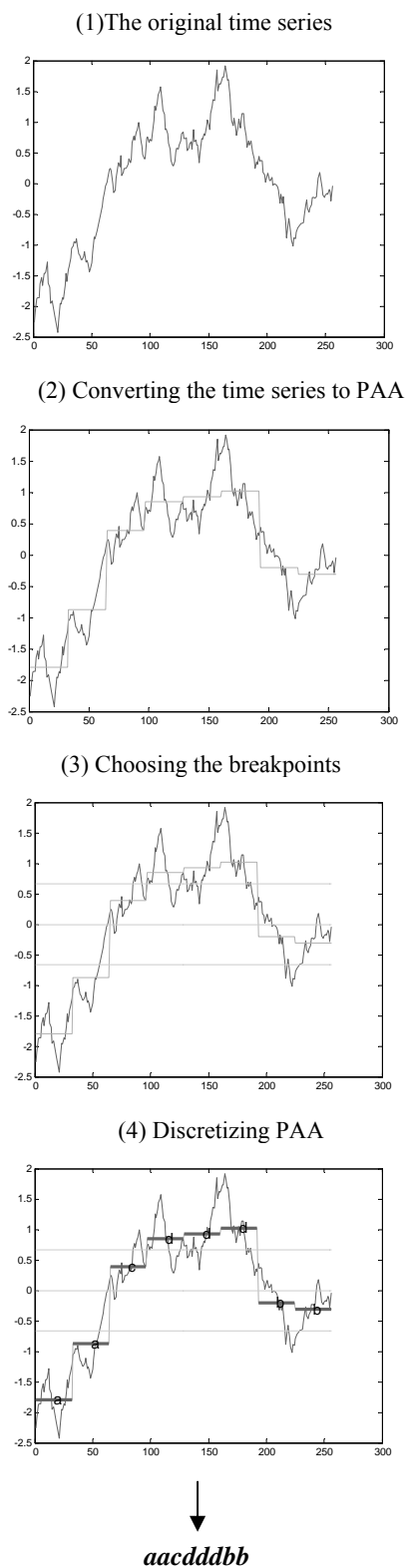

(1) The original time series

(2) Converting the time series to PAA

(3) Choosing the breakpoints

(4) Discretizing PAA

*aacdddbb*

Figure 1: The different steps of applying SAX

representations of the two time series $S$ and $T$, respectively, and where the function $dist(\ )$ is implemented by using the appropriate lookup table.

We need to mention that the similarity distance used in PAA is:

$$d(X,Y) = \sqrt{\frac{n}{N}} \sqrt{\sum_{i=1}^{N} (\bar{x} - \bar{y})^2} \qquad (4)$$

Where $n$ is the length of the time series, $N$ is the number of frames. It is proven in (Keogh, et al. 2000) and (Yi and Faloutsos 2000) that the above similarity distance is a lower bounding of the Euclidean distance applied in the original space of time series. This results in the fact that MINDIST is also a lower bounding of the Euclidean distance, because it is a lower bounding of the similarity distance used in PAA. This guarantees no false dismissals. Figure.1 illustrates the different steps of SAX.

## 3  THE PROPSED METHOD

The basis of our method is to improve the performance of SAX by combining it with an exclusion condition that increases its pruning power, and by using different reduced spaces, and not only one.

We divide each time series into $N$ segments. Each segment is approximated by a polynomial. In this paper we use a polynomial of the first-degree for its simplicity, but other approximating functions can be used as well.

Since this approximating function is the optimal approximation of the corresponding segment, the distance between this segment and this approximating function is minimal. The time series is also represented using SAX in the way indicated in section 2.2. A polynomial of the same degree is used to approximate all the segments of all the time series in the database. So now each time series has two representations: the first is a $n$-dimensional one, by using the approximating function, and the second is a $N$-dimensional one, by using SAX. We also have two similarity distances; the first one is the Euclidean distance, which is the distance between a time series and its approximating function, and the other one is MINDIST, which is the distance in the reduced space.

Given a query $(q,\varepsilon)$, let $\bar{u}$, $\bar{q}$ be the projections of the $u$, $q$, respectively, on their approximating functions, where $u$ is a time series in the database.

By applying the triangular inequality we get:

$$d(\bar{q},u) \leq d(q,u) + d(q,\bar{q}) \qquad (5)$$

Taking into consideration that $\bar{u}$ is the best approximation of $u$, we get:

$$d(u,\bar{u}) \leq d(u,\bar{q}) \qquad (6)$$

By substituting the above relation in (5), we can safely exclude all the times series that satisfy:

$$d(u,\bar{u}) > \varepsilon + d(q,\bar{q}) \qquad (7)$$

In a similar manner, and since $\bar{q}$ is the best approximation of $q$, we can safely exclude all the time series that satisfy:

$$d(q,\bar{q}) > \varepsilon + d(u,\bar{u}) \qquad (8)$$

Both (7), (8) can be expressed in one relation:

$$\left| d(u,\bar{u}) - d(q,\bar{q}) \right| > \varepsilon \qquad (9)$$

In addition to the exclusion condition in (9), since MINDIST is lower bounding of the original Euclidean distance, all the time series that satisfy:

$$MINDIST(\bar{q},\bar{u}) > \varepsilon \qquad (10)$$

Should also be excluded, Relation (10) defines the other exclusion condition.

**The Offline Phase:** The application of our method starts by choosing the lengths of segments. We associate each length with a level of representation. The shortest lengths correspond to the lowest level, and the longest lengths with the highest levels. Each series in the database is represented by a first-degree polynomial, which is the same approximating function for all the time series in the databases. The distances between the time series and their approximating function are computed and stored. In order to represent the time series, we choose the alphabet size to be used. SAX appeared in two versions; in the first one the alphabet size varied in the interval (3:10), and in the second one the alphabet size varied in the interval (3:20). We choose the appropriate alphabet size for this datasets. The time series in the database are represented using SAX on every representation level

**The Online Phase:** The range query is represented using the same scheme that was used to represent the time series in the database. We start with the lowest level and try to exclude the first time series using (9) if this time series is excluded, we move to the next time series, if not, we try to exclude this time series using relation (10). If all the time series in the database have been excluded the algorithm terminates, if not, we move to a higher level. Finally, after all levels have been exploited, we get a potential answer set, which is linearly scanned to filter out all the false alarms and get the true answer set.

## 4 EXPERIMENTS

We conducted experiments on different datasets available at UCR (UCR Time Series datasets) and for all alphabet sizes, which vary between 3 (the least possible size that was used to test the original SAX) to 20 (the largest possible alphabet size). We compared the speed of our method FAST_SAX, with that of SAX as a standalone method. The comparison was based on the number of operations that each method uses to perform the similarity search query. Since different operations take different execution times, we used the concept of *latency time* (Schulte et al. 2005). We report in Tables 1 and in Figure 2 the results of (wafer). We chose to present the results of this dataset because it is the largest dataset in the repository. Also it is shown in (Muhammad Fuad and Marteau 2008) that the best results obtained with SAX were with this dataset.

The codes of SAX were optimized versions of the original codes, since the original codes were not optimized for speed

Table 1: Comparison of the latency time between SAX and FAST_SAX for ε=1:4 and alphabet size=3, 10, 20

|          | α= 3      | α= 10     | α= 20     |
|----------|-----------|-----------|-----------|
| FAST_SAX | 1.0592E6  | 3.7136E5  | 2.6734E5  |
| SAX      | 5.1291E6  | 1.5764E6  | 1.2759E6  |

ε=1
(a)

|          | α= 3      | α= 10     | α= 20     |
|----------|-----------|-----------|-----------|
| FAST_SAX | 7.2062E6  | 3.3509E6  | 2.2255E6  |
| SAX      | 1.567E7   | 4.8078E6  | 3.4253E6  |

ε=2
(b)

|          | α= 3      | α= 10     | α= 20     |
|----------|-----------|-----------|-----------|
| FAST_SAX | 1.3717E7  | 1.1444E7  | 9.5446E6  |
| SAX      | 2.1944E7  | 1.4428E7  | 1.1144E7  |

ε=3
(c)

|          | α= 3      | α= 10     | α= 20     |
|----------|-----------|-----------|-----------|
| FAST_SAX | 2.2697E7  | 1.6928E7  | 1.6611E7  |
| SAX      | 2.8287E7  | 2.2179E7  | 1.9877E7  |

ε=4
(d)

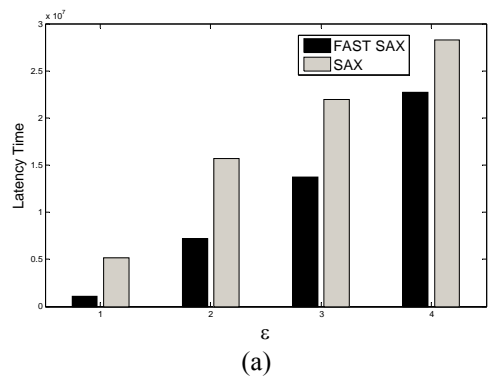

(a)

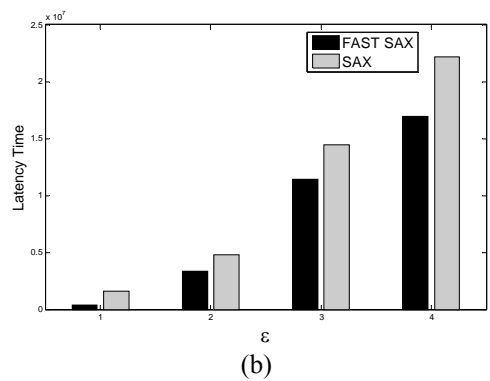

(b)

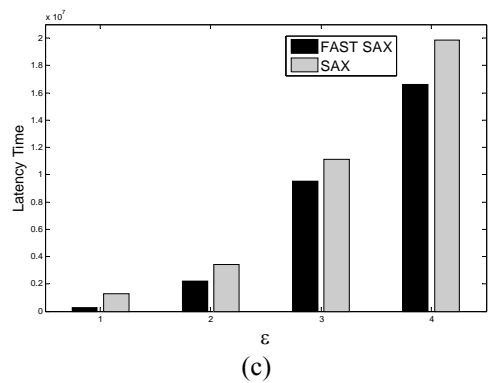

(c)

Figure 2: Comparison of the latency time between SAX and FAST_SAX for alphabet size=3 (a), alphabet size=10 (b) and alphabet size=20 (c)

The results of testing our method on other datasets are similar to the results obtained with (wafer)

The results shown here are for alphabet size 3 (the smallest alphabet size), 10 (the largest alphabet size in the first version of SAX), and 20 (the largest alphabet size in the second version of SAX)

The results obtained show that FAST_SAX outperforms SAX for the different values of ε and for the different values of the alphabet size.

## 5 CONCLUSION AND PERSPECTIVES

In this paper we presented a method that speeds up the symbolic aggregate approximation (SAX). We conducted several experiments of times series similarity search, with different values of the alphabet size and different threshold values. The results obtained show that the new method is faster than the original SAX

The future work can focus on extending this method to other representation methods, other than SAX, also on using other approximating functions.